# The Search for Dark Matter (WIMPS) at Low Mass

David B. Cline[*]
UCLA Physics & Astronomy
Astroparticle Physics Group

**Abstract**

We briefly review the constraints on the search for low mass wimps (< 15 GeV) and the various experimental methods. These experiments depend on the response of detectors to low energy signals (less than 15 KeV equivalent energy). We then describe recent fits to the data and attempt to determine $L_{eff}$, the energy response at low energy. We find that the use of a liquid Xenon 2-phase detector that employs the $S_2$ data near threshold is the most sensitive current study of the low mass region. We rely on some talks at Dark Matter 2010.

## 1. Introduction

The search for particle dark matter is among the most difficult of current experimental efforts. Some dark matter makes up about one quarter of the matter in the universe, is a key element in the formation of galaxies, and could well lead to new understandings in elementary particle physics. The detection of dark matter will be of very great significance in science.

UCLA has hosted ten workshops on dark matter detection, the most recent taking place in February 2010. While there have been several key developments since then, these reports are likely a state of the art source for direct and indirect detection of dark matter. These talks can be found at:

http://www.physics.ucla.edu/hep/dm10/presentations.html

In this report we cover the issue of low mass (below 15 GeV) WIMP detection. There is no real theoretical motivation for such low mass particles, but we don't have a theory for dark matter in general either, so the search must go on.

## 2. Current state of world data

The DAMA/LIBRA results have been extensively studied by G. Gelmini [1] and colleagues. They have fit all the data and find two WIMP mass regions as shown in Figure 1. We study here the low mass region (~10GeV) which has a lower probability in their fit. The higher mass solution is ruled out by a large factor by 10 or more by direct dark matter search experiments. [5 experiments] [see references 2,3].

---

[*] This work carried out at the Aspen Center for Physics.

These results and the possible WIMP signal from the CoGeNT experiment along with the current constraints taken on face value from CDMSII, XENON10, and XENON100 data are shown in Figure 2 [4]. The constraints on the low mass region (~7GeV/c²) have been called into question recently. We review the studies in this region in this article.

Now three experiments have claimed a signal in the low wimp mass region: DAMA, CoGeNT and now CRESST. The analysis of 2-phase detector data near threshold is even more important. We report on these (new) data here.

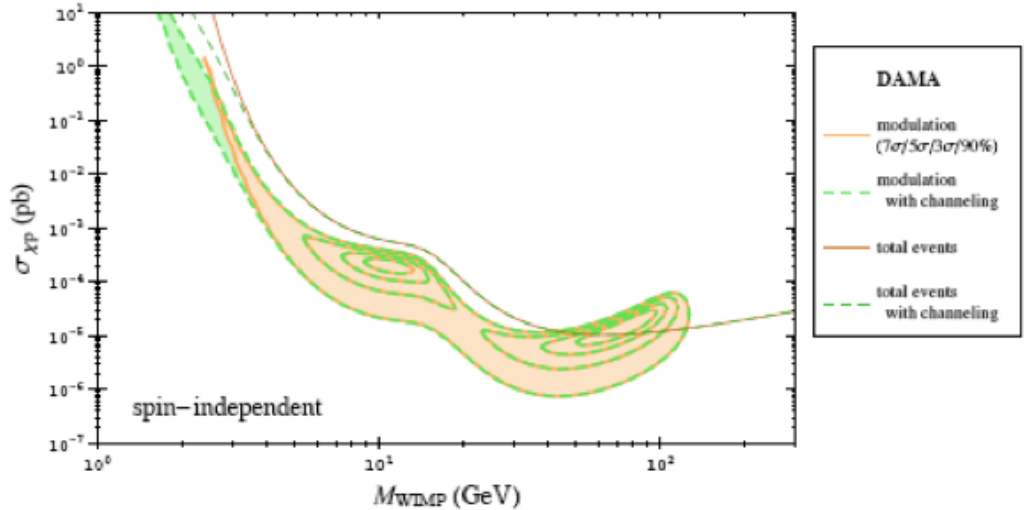

Figure 1. WIMP masses and spin-independent (SI) cross-sections compatible with the DAMA modulation signal and total number of events, determined with (dashed green) and without (solid orange) the channeling effect included. The largest channeling fractions shown in Figure 1 (taken from Ref. [3]) are used here for the channeling case. Comparing the cases with or without channeling, we find negligible difference in the DAMA modulation regions at the 90%, 3σ, and 5σ levels; only the 7σ contours differ and only for WIMP masses below 4 GeV. The lower and higher mass DAMA regions correspond to parameters where the modulation signals arise from scattering predominantly off of NA and I, from Reference 2.

CoGeNT and DAMA (from Reference 4 for references)

1) The CoGeNT experiment uses a small Ge crystal to search for a coherent interaction than just experiment used reaction neutrinos to search for antineutrino coherent scattering. They made impressive gains but the low energy background obscured the effect. The detector was sent underground to Soudon and a search was made of coherent WIMP scattering. A similar signal on background was observed (see Juan Collar's talk at DM 2010 website).
2) The DAMA /Libra experiment is well known. Reference 4 gives many of the original references.

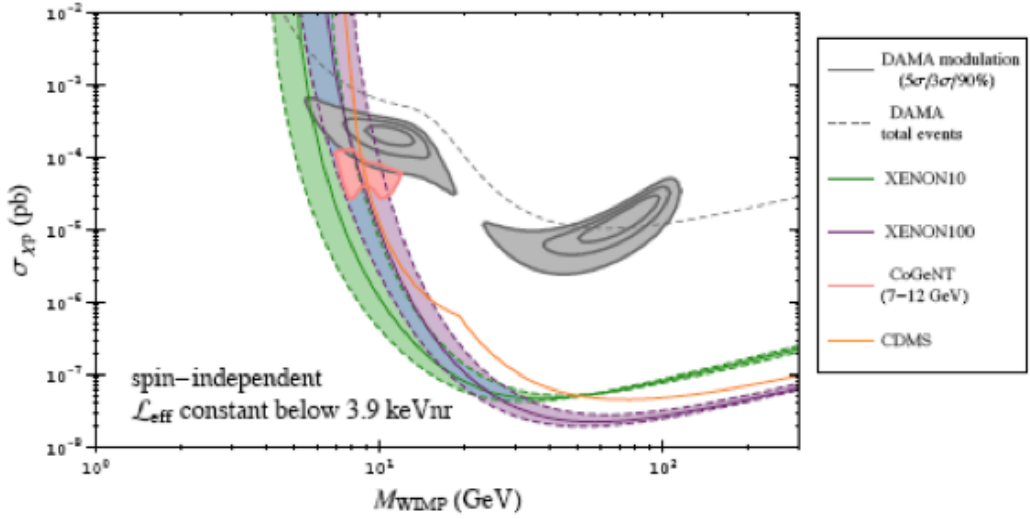

Figure 2. XENON10 (green) and XENON100 (purple) 90% C.L. constraints for a constant $\mathcal{L}_{eff}$ at recoil energies below 3.9 KeVnr. The solid curves are the constraints using the central values of $\mathcal{L}_{eff}$ as described in the text; dashed curves and lighter filled regions indicate how these 90% constraints vary with the 1σ uncertainties in $\mathcal{L}_{eff}$. The blue region indicates an overlap between the XENON10 (green) and XENON100 (purple) 1σ regions. Also shown are the CDMS constraint (orange curve), DAMA modulation compatible regions (gray contours/region), and the CoGeNT 7-12 GeV region (pink contour/region). The lower and higher mass DAMA regions correspond to parameters where the modulation signals arise from scattering predominantly off of Na and I, from Reference 4 see also Reference 3..

Recently the XENON100 experiment has started taking data. While a blinded search is underway the collaboration decided to publish the results from eleven days of unblinded data (to be published in Physics Review Letters).

In the course of this analysis it was understood that the poisson distribution of the data for the low mass (and low energy transfer) along with the absence of any candidate events leads to the exclusion plot in Figure 3. This plot is dependent on the detector response at low energy transfer that we discuss in this report. Again the experiment excludes the low mass region of DAMA/LIBRA and CoGeNT (Figure 3, see also Reference 4).

We only show the three CRESST events from DM 2010. In Figure 4 we show the recent state of the dark matter search from the CRESST detector. Note that they had three candidates for dark matter at that time (February 2010). The XENON 100 data appears to rule out these results. These data were reported at the Marina del Rey Dark Matter experiment and the talk can be found in the talks referenced before.

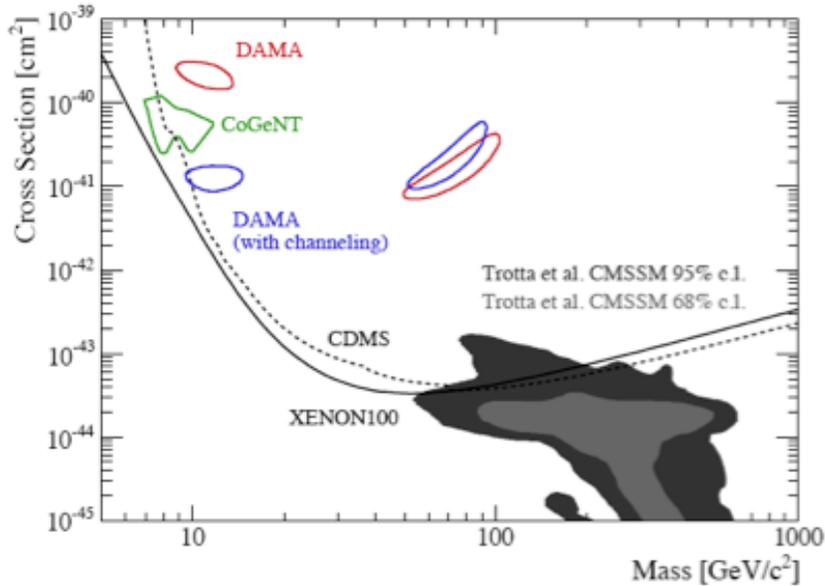

Figure 3. 90% confidence limit on the spin-independent elastic WIMP-nucleon cross-section (solid line), together with the best limit to date from CDMS (dashed), expectations from a theoretical model, and the areas (90% CL) favored by CoGeNT (green) and DAMA (blue/red). [Taken from Ref. 2].

The CRESST experiment is a cryogenic detector located at the Gran Sasso underground laboratory (see the talk of L. Stodolsky at DM 2010). The detector has recently been upgraded.

At the Dark Matter 2010 meeting the first data from the upgraded detector was shown (you can find this on the website for DM 2010 previously discussed). We show this plot in Figure 4. There are three events in the dark matter band. If these events were due to WIMP interaction they would be consistent with the DAMA and CoGeNT results (Figure 2). Since no paper has been reported on the possible signal we will not discuss this further in this paper.

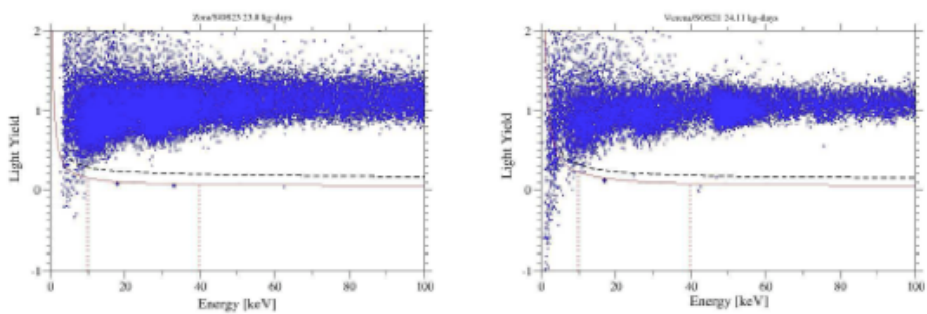

Figure 4. CRESST candidate events shown at Dark Matter 2010 at UCLA. The events are in the lower bonds (unpublished).

## 3. Interactions of low mass wimps with matter

The mean velocity of WIMPS in our galaxy is 250km/sec. The recoil energy of the WIMP, $m_w$, is the mass of the WIMP. $m_A$ is the mass of the target nucleus. The galactic velocity is [250km/sec/3 x $10^5$km/sec]x c and target range from 0 to $m_A$. If the WIMP mass is small formula 1 gives for the recoil energy [5]

$$E_v \sim 4 E_0 \frac{m_w}{m_t}$$

Note: lower mass targets yield larger $E_v$. However if $m_w$ is very small then the recoil energy is again reduced. So low mass WIMPS even with low mass targets yield small recoil energy. For example, for oxygen target A=16 and wimp mass of 5 GeV we get the recoil energy of

$$E_v = \left(\frac{20}{16}\right) E_0$$

In many current experiments $m_t$ is large and there are more sensitive to high mass WIMPs of 100 GeV or so. However this is not the energy that will be detected in reality since two factors must be included:

(1) The form factor of the recoiling nucleus
(2) The "quenching" factor for the nucleus and material of the detector.

The quenching factor is the ratio of

$$Q = \frac{observed\ energy}{E_v}$$

Q depends on the method of detection (i.e. cryogenic detectors or ionization detectors) and typically runs from 0.1 to 0.3 in most experiments.
Quenching factors (or the equivalent) can be measured using incident neutron beams since the recoil energy should be the same as that of a wimp interaction with the same $E_0$. The quenching factor can be calculated by the so-called Lindhard method. This gives an approximate quenching factor and is unreliable for low mass WIMPs and low energy recoils. This directly relates to the difficulty in detecting low mass WIMPs which would give low energy recoils. This directly relates to the difficulty in detecting low mass WIMPs that would give low energy recoils (see Reference 5).
As we will see there are three methods to attempt to resolve this difficulty:

(1) Improve low energy recoil experiments using neutron beams

(2) Use of the $S_2$ signal in two-phase experiments as an internal calibration method
(3) Estimate the effects of the uncertainty of the recoil energy response on the detection signal, especially if comparing different experimental results.

## 4. The use of the $S_2/S_1$ and $S_2$ methods in two-phase detectors for the low mass regions

From the start of the concept of the $S_2/S_1$ discrimination in the 1990s, we knew that there would be problems when $S_1$ got very small or near threshold for the design of ZEPLIN II [6]. An internal analysis was made by DBC of the ZEPLIN II data using only $S_2$ data (when $S_1$ was very small). This turned out not to help the sensitivity for ZEPLIN II because of the large PMTs which did not allow a precise fiducial volume to be defined and was never published. However for future analyses the $S_2$ method may be useful.

While it may seem that the low mass WIMP region should be probed by lower mass targets (like oxygen or sodium), this is not the case if the two-phase method is used. In this method electrons generated by the recoil nucleus moving through the medium generate free electrons. The application of an electrical field on the detector is used to drift the free electrons into a volume of gas with a higher electrical field that provides an amplification of the signal. This can overcome any advantage that low mass might offer. Very approximately for every free electron produced at the WIMP interaction vertex there will be a 25 $P_e$ ($P_e$ photoelectric) in the gas phase.

This concept was invented to be used for liquid Xenon detectors at UCLA and Torino universities in the mid-1990s but is now being used for liquid Argon detectors (WARP, Ar or Ne, etc.) and others [6]. In this paper we only cover the use in liquid Xenon detectors with published data.

## 5. The use of the $S_2$ signal

The key to the use of $S_2$ is to define a small fiducial volume. X and Y positions are measured directly in the PMT structure. The determination of the Z point of the event uses the time spread of the $S_2$ signal produced at different depths in the detector, since the X,Y signal can define a small fiducial volume even in a poorly measured Z position and maintain this small volume.

As stated before, an internal study using $S_2$ data was carried out for the ZEPLIN II data at UCLA. No analysis was done with the data due to the large backgrounds. More recently a very nice study of the use of $S_2$ data to determine the sensitivity of XENON10 data to low mass WIMPs was done by P. Sorensen [7]. We repeat some of this analysis here for completeness.

## 6. Analysis of the sensitivity of the limit on low mass WIMPs due to threshold effect

Another approach first carried out by the XENON100 team is the two different $\mathcal{L}_{\text{eff}}$ behaviors near threshold based on data and then determine the sensitivity of the limit (see Fig. 7). Recently Savage et al [4] have carried out a more complete study.

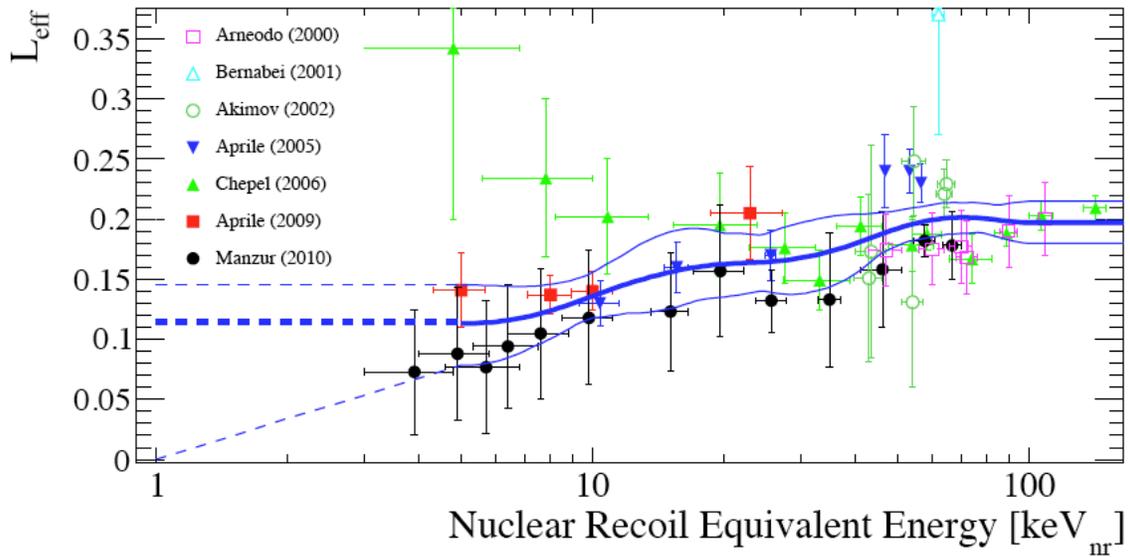

Figure 5. From Reference 2, showing the fits for $\mathcal{L}_{\text{eff}}$ by the XENON100 team.

$\mathcal{L}_{\text{eff}}$ is defined as:

$$\mathcal{L}_{\text{eff}} = \frac{S_0 S_e}{E_\gamma L_y S_n}$$

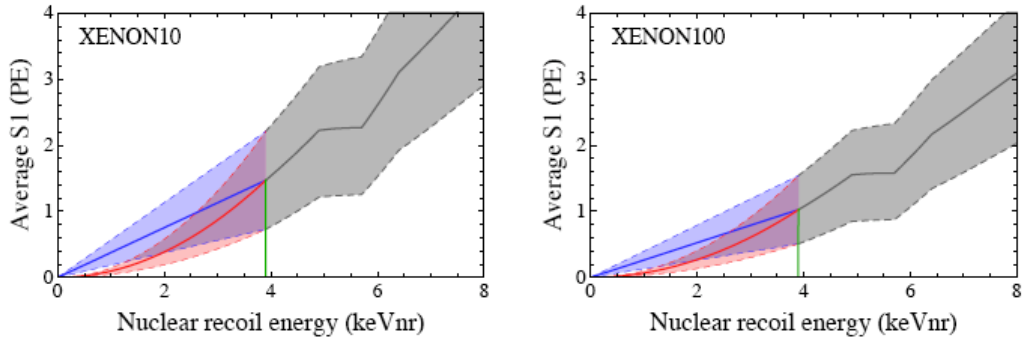

Figure 6. Study of Savage et al on $\mathcal{L}_{\text{eff}}$ from Reference 4.

The behavior of the signal close to threshold can be calculated at present. However the standard model would apply to all targets at some level so it is likely that the response sort of scales, i.e. is the quenching factor of I is 0.1. We would expect a similar result for other heavy nucleons like W or same is likely to be true for $X_e$ (Xenon).

**7. Exclusion limits for low mass wimps**

There is considerable controversy on the WIMP mass region that is excluded by the XENON10 and XENON100 experiments. There are new and unconfirmed claims of more candidate events from the CRESST experiment. Clearly new measurements are needed.
   (1) The XENON100 experiment is currently taking data that is "blinded," namely the data will not be studied until the "box is opened" later this year. These data should be at least a factor of 5 more sensitive than current data.
   (2) New direct measurements of the quenching factor or $\mathcal{L}_{\text{eff}}$ need to be carried out. There are plans in the XENON100 group to carry out these measurements in the next year, we understand.

However while waiting for these new measurements we can draw some tentative conclusions from the preceding arguments about:

   (1) The effect of uncertainty in $\mathcal{L}_{\text{eff}}$
   (2) The use of the $S_2$ method to probe the sensitivity of the Xenon detectors to signals very close to threshold

The work by Savage et al (Refernece 4) and the XENON100 paper both address issue 1.

In the now published XENON100 paper that uses 11 days of "unblinded" data the authors state "as shown in Figure 3 over the acceptance is sizeable even at a reduced threshold of 3PE (8.2 KeV in this case) above which a 7 GeV/$c_2$ WIMP at the lower edge of the CoGeNT region would produce about one event with the current exposure."

In this case the forthcoming new XENON100 data would have between 5-10 events in this case. So the new data should resolve the issue.

The paper of savage et al takes a different approach to this issue as shown in Figure 2. They analyze both XENON10 and XENON100 data for the first time. They also try different forms of $\mathcal{L}_{\text{eff}}$ near threshold (Figure 9). When the first analysis is done they find an exclusion plot shown in Figure 7.

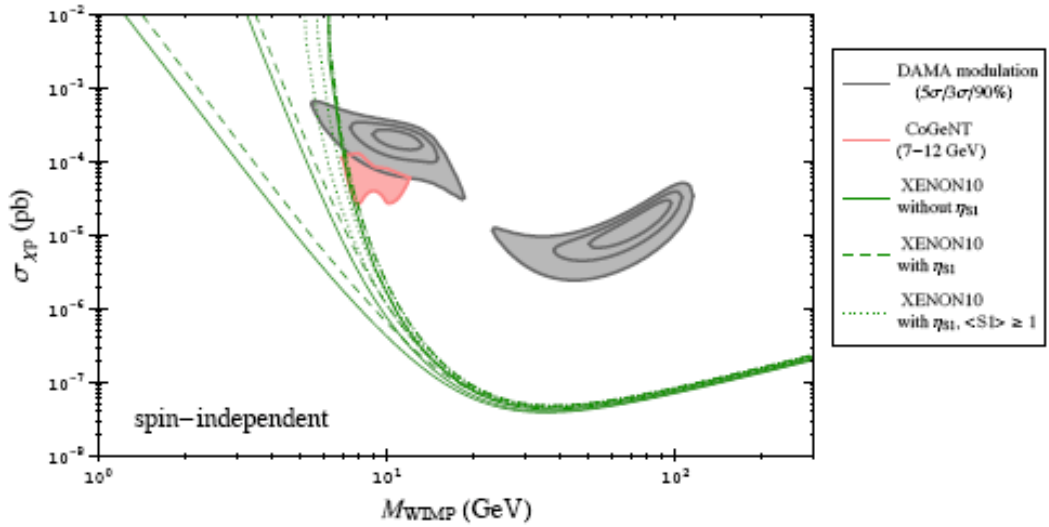

Figure 7 (Reference 4). Exclusion plot of Savage et al.

The use of the $S_2$ method to determine the near threshold behavior of $\mathcal{L}_{eff}$ (or equivalent) was described in Reference 7. We show the results of this analysis in Figure 8.

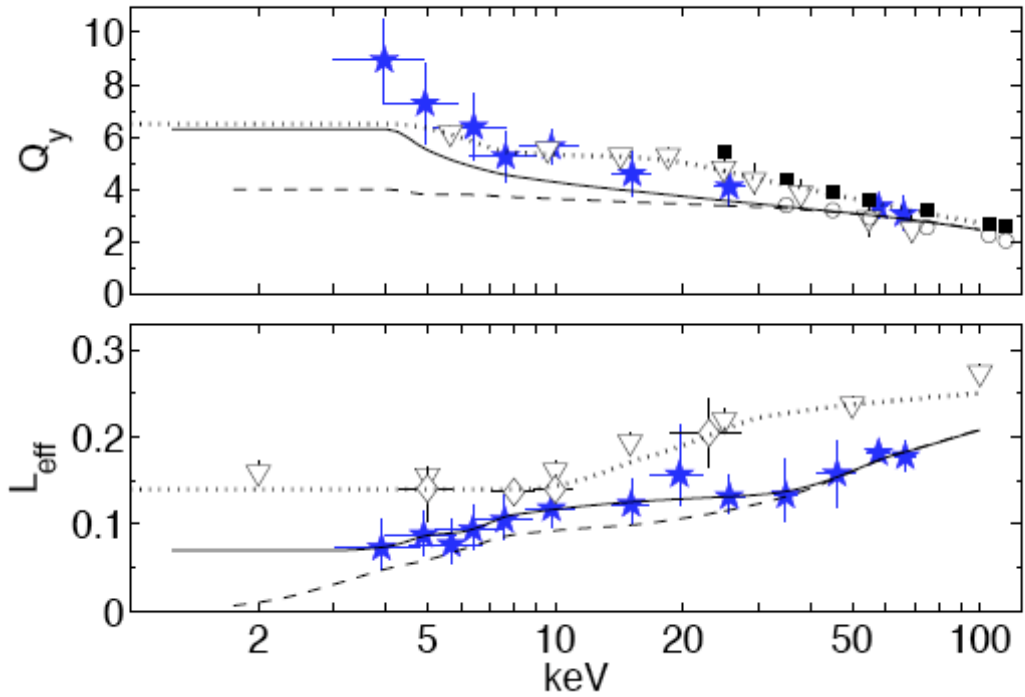

Figure 8. Study of $\mathcal{L}_{eff}$ by Sorensen using $S_2$ signal data (see Reference 7).

It is clear that the more robust signal near threshold is preferred by the data. Using this analysis results in an exclusion plot of Figure 9.

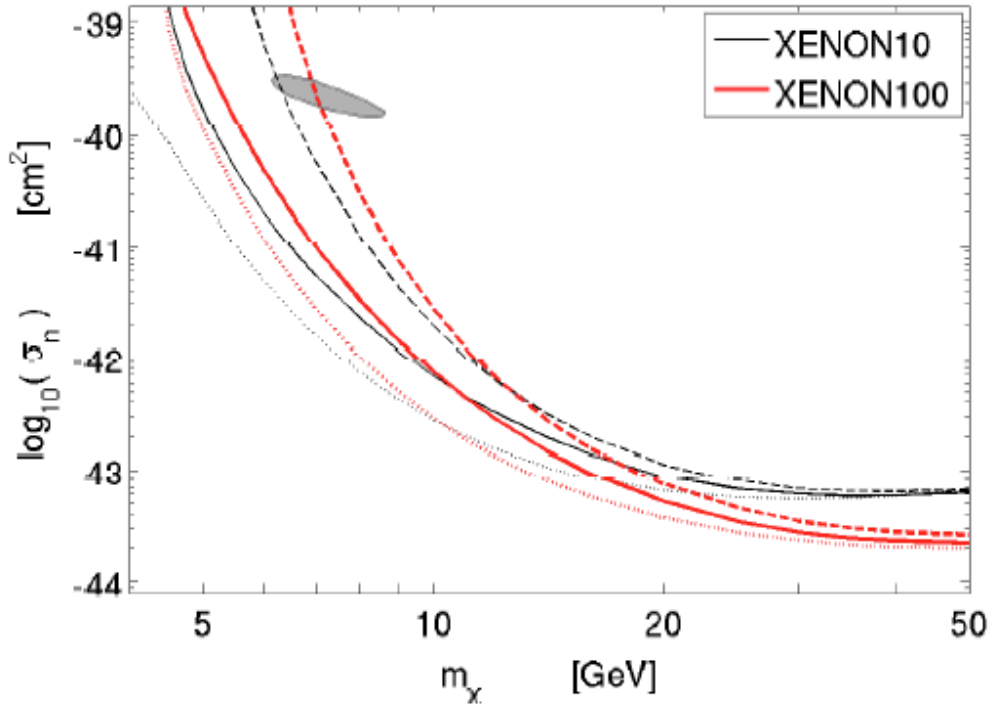

Figure 9. Exclusion plot by Sorensen using the $S_2$ determined $\mathcal{L}_{\text{eff}}$ [7].

The XENON100, Savage et al and the $S_2$ analysis of Sorensen all show a similar pattern of exclusion of the low mass WIMPs. The results of the exclusion plot using this method is shown in Figure 9 [Reference 7].

Summary

The current situation in the search for low mass WIMPs is confused. Two experiments give a hint of a signal. However the XENON10 and XENON100 (and CDMS II) seem to rule out much of the phase space and the analyses of Savage et al (Reference 4). Sorensen (Reference 7) is very promising using the $S_2$ signal from the XENON two-phase detectors (XENON10).

I wish to thank Danny Marfatia of Kansas University for help with this note and the XENON100 group that are not responsible for this summary, as well as Katherine Freese, Graciela Gelmini, and Elena Aprile for long-time discussions.

References

1. G. Gelmini and P. Gondolo, arXiv:hep-ph/0405278; P. Gondolo and G. Gelmini, Phys. Rev. D 71, 123520 (205) [arXiv:hep-ph/0504010].